\documentclass[12pt,fleqn]{article}
\usepackage[]{graphicx}
\begin{document}
\title{SPECTRAL STRUCTURE OF SINGULAR SPECTRUM DECOMPOSITION FOR TIME SERIES}

\author{Kenji Kume \thanks{Department of Physics, Nara Women's University, Nara 630-8506, Japan. } \and Naoko Nose-Togawa \thanks{Research Center for Nuclear Physics, Osaka University, Ibaraki 567-0047, Japan}  }

\maketitle
\thispagestyle{empty}

\begin{abstract}
Singular spectrum analysis (SSA) is a nonparametric and adaptive spectral decomposition of  a time series. The singular value decomposition of the trajectory matrix and the anti-diagonal averaging leads to a time-series decomposition. 
In this algorithm, a single free parameter, window length $K$, is involved which is the FIR filter length for the time series. There are no generally accepted criterion for the proper choice of the window length $K$. Moreover, the proper window length depends on the specific problem which we are interested in. Thus, it is important to monitor the spectral structure of the SSA decomposition and its window length dependence in detail for the practical application.  In this paper, based on the filtering interpretation of SSA, it is shown that the decomposition of the power spectrum for the original time series is possible with the filters constructed from the eigenvectors of the lagged-covariance matrix. With this, we can obtain insights into the spectral structure of the SSA decomposition and it helps us for the proper choice of the window length in the practical application of SSA.
\end{abstract}

\newpage

\section{Introduction}
\label{intro}
Singular Spectrum Analysis (SSA) is a nonparametric and adaptive decomposition of a time series.  In the SSA, the original time series is separated into arbitrary number of additive components which are grouped and are interpreted as the slowly varying trend, oscillatory and noise components, respectively.  It has been applied to many different fields from physics and earth sciences to financial and social sciences  [2-7]. 

In the conventional treatment of SSA, the algorithm consists of four steps:
\begin{itemize}
\item 1st step: a time series  $\{ x_0,x_1,x_2,...,x_{N-1} \} $ of length $N$ is transformed into multidimensional series by constructing the $L \times K$ trajectory matrix $X$.  Where $K$ is the window length which is an arbitrary positive integer satisfying $K \leq N/2$ and $L=N-K+1$.
\item 2nd step:  The singular value decomposition of the trajectory matrix
\begin{equation}
X=X_1+X_2+...+X_K \ \ \ .
\label{Eq1}
\end{equation}  
\item 3rd step:  so-called grouping procedure where the above set of rank-one matrices $X_k$ is partitioned into several groups. 
\item 4th step: reconstruction process where the decomposition of the time series is completed with the anti-diagonal averaging of the grouped matrix $X^{(j)}=X_{i_1}+X_{i_2}+X_{i_3}+...+X_{i_m}$ which transform $X^{(j)}$ into Hankel form.
\end{itemize}

The major objective of the SSA algorithm is to separate the original time series into interpretable set of time series.  A single parameter, window length $K$, is involved in the algorithm and the results of decomposition depend on the choice of this parameter $K$.   These points are throughly discussed from the {\it separability} point of view
[3,4].  It is argued that the relatively large window lengths are preferable to achieve a better (weak) separation, but for the series with a complex structure, too large window length leads to undesirable decomposition or mixture of the series components.  Then, how to choose the proper window length $K$ is an important point for the practical application of SSA but is not easy to formulate in a general way. The choice of the window lengths have been done so far by investigating the behavior of the eigentriples (singular values, and left and right singular vectors of the trajectory matrix) or the w-correlation matrix.

 In this paper, based on the filtering interpretation of SSA [1,8,10,12], the decomposition of the power spectrum for the original time series is shown to be possible and this gives us detailed information about the spectral structure of the SSA decomposition. 
In the filtering interpretation, the SSA algorithm can be seen as the generation of normalized and complete set of adaptive filters and the operation of these filters to the original data.  The SSA decomposition is thus interpreted as the soft and the optimal partitioning of the Fourier space through the adaptively constructed filters. By utilizing the completeness property of the eigenvectors of the lagged-covariance matrix, the  filters constructed with these eigenvectors are used for the decomposition of the power spectrum of an original time series.  The decomposition of the power spectrum shows the detailed structure of the SSA decomposition and, with this, we can obtain useful  
 information about the spectral structure of the SSA decomposition and this helps us to choose the proper window length $K$ in the practical application. 

In the next Sec. 2, we briefly summarize the conventional treatment of the SSA algorithm. We also describe the filtering interpretation of the SSA and it is adopted for the decomposition of the power spectrum of original time series.  In Sec. 3, in order to see how the decomposition of the power spectrum works, we have examined two cases: the schematic time series with the superposition of the pure harmonic components and the daily  
 currency exchange rates of EUR/USD. The summary and conclusions are given in Sec. 4.

\section{Algorithm for the singular spectrum analysis}
\label{sec:1}
In this section, we briefly describe the SSA algorithm.  In the filtering interpretation, the SSA algorithm consists of two steps. First, optimal and adaptive construction of the normalized and complete set of filters through the eigenvectors of the lagged-covariance matrix. Second step is the operation of these filters to the original time series. In the conventional treatment of the SSA, these two procedures are mixed into the singular value decomposition and the anti-diagonal averaging procedure.  Because of the completeness property of the filters generated in SSA, not only the time series but also the power spectrum can be decomposed into arbitrary number of {\it additive} components.  In order to show this, we assume the periodic boundary condition for the time series to avoid subtleties coming from the starting or the end points of a time series.    
\subsection{Conventional SSA algorithm}
\label{sec:2}
Let us consider a real-valued time series $\{ x_0,x_1,x_2,...,x_{N-1} \}$ of sufficient length. First, we define the $L \times K$ trajectory matrix $X$ as
\begin{equation}
X_{ij}=x_{i+j-2} \ \ \ \ \ (i=1,2,...,L; \ j=1,2,...,K) \ \ \ \ ,
\label{Eq2}
\end{equation}
where window length $K$ is the arbitrary positive integer satisfying $K \leq N/2$ and $L=N-K+1$.
Next step is the singular value decomposition of the above trajectory matrix $X$. 
To do this, the lagged-covariance matrix $C=X^TX$ is introduced and the eigenvalue equation is solved
\begin{equation}
C{\bf v}^{(k)}=\lambda_k{\bf v}^{(k)} \ \ \ .
\label{Eq3}
\end{equation}
The eigenvalues $\lambda_k$ are ordered as the decreasing order of magnitude $(\lambda_1,\lambda_2,...,\lambda_K)$ and the corresponding eigenvectors are denoted as ${\bf v}^{(k)}$. We define the $L$ dimensional vector as
${\bf w}^{(k)}=X{\bf v}^{(k)}/\sqrt{\lambda_k}$. Then the $L \times K$ rank-one matrices $X_k$ are expressed as $X_k=\sqrt{\lambda_k}{\bf w}^{(k)}{\bf v}^{(k)T}$. Then the singular value decomposition of the trajectory matrix is expressed as
\begin{equation}
X=\sum_{k=1}^K X_k= \sum_{k=1}^K \sqrt{\lambda_k}{\bf w}^{(k)}{\bf v}^{(k)T} \ \ \ .
\label{Eq4}
\end{equation}
The third step is the partitioning of the matrices $X_k$ into several groups and is called as the grouping procedure. The fourth step is the reconstruction of the time series. For the grouped matrix $X^{(j)}=X_{i_1}+X_{i_2}+X_{i_3}+...+X_{i_m}$, we carry out the anti-diagonal average which transform $X^{(j)}$ into Hankel form. Then the time series are reconstructed giving the SSA decomposition. 

\subsection{Filtering interpretation of SSA algorithm}
 As are discussed in Refs. [8,10-12], above mentioned SSA algorithm can be viewed as the two-step point-symmetric filtering of the time series. The filtering interpretation is much more transparent in Fourier space, and the normalization and the 
 completeness properties of the filters play a crucial role in SSA decomposition.
To see this, we define the discrete Fourier transform as 
\begin{equation}
\hat{x}_\alpha = {1 \over \sqrt{L}} \sum_{j=0}^{L-1} \rm{exp}(2 \pi i \it{}\alpha j/L) \ x_j \ \  , \label{Eq5}
\end{equation}
and
\begin{equation}
\hat{v}_\alpha^{(k)} = \sum_{j=0}^{K-1} \rm{exp}(2 \pi i \it{}\alpha j/L) \ v_{j+\rm{1}}^{(k)} \ \ , 
\label{Eq6}    
\end{equation}
where the components of the vector ${\bf v}^{(k)}$ is denoted as $(v_1^{(k)},v_2^{(k)},...,v_K^{(k)})^T$.  In above Eq.(\ref{Eq6}), we consider the Fourier transform of the series of length $L$ $(v^{(k)}_1, v^{(k)}_2,...,v^{(k)}_K,0,0,...,0)$ and then the sum $\sum_{j=0}^{L-1}$ is replaced by $\sum_{j=0}^{K-1}$.

The forward filtering $X{\bf v}$ can be rewritten in the Fourier space as
\begin{equation}
\hat{x}_\alpha \longrightarrow \ \ \ \hat{v}_{-\alpha}^{(k)} \hat{x}_\alpha =  \hat{v}_{\alpha}^{(k)*} \hat{x}_\alpha\ \ .
\label{Eq7}
\end{equation}
The above relation holds for the periodic series. Since the eigenvectors $v^{(k)}_j$ are real valued, their Fourier transform satisfies  $\hat{v}^{(k)}_{-\alpha}=\hat{v}^{(k)*}_{\alpha}$.  In above equation, the vectors $\hat{v}_\alpha^{(k)}$ are complex in general, and shift and distort the time series. The non-causal FIR filtering Eq. (\ref{Eq7}) corresponds to the SVD of the trajectory matrix in the conventional treatment of SSA. The last step of SSA is the anti-diagonal averaging (Hankelization) for the rank-one matrix $X_k$ and this can be expressed as the causal FIR filtering 
\begin{equation}\ \ \ .
\hat{x}_\alpha \longrightarrow \hat{v}_\alpha^{(k)} \hat{x}_\alpha \ \ .
\label{Eq8}
\end{equation}
Then the two-step filtering can be combined as a single real filtering 
\begin{equation}
\hat{x}_\alpha \longrightarrow |\hat{v}_\alpha^{(k)}|^2 \hat{x}_\alpha \ \ \ ,
\label{Eq9}
\end{equation}
and does not distort the time series [1,9,10,12].
It should be important to note that the orthonormal and the completeness properties of the eigenvector ${\bf v}^{(k)}$ lead to the following relations for the filters $ |\hat{v}_\alpha^{(k)}|^2$ [10,11]
\begin{equation}
\sum_{\alpha=0}^{L-1} |\hat{v}^{(k)}_\alpha|^2/L=1 \ \ \ ,
\label{Eq10}
\end{equation} 
and 
\begin{equation}
\sum_{k=1}^{K} |\hat{v}^{(k)}_\alpha|^2/K=1 \ \ \ .
\label{Eq11}
\end{equation}
An arbitrary orthonormal $K$-dimensional vector set $\{  {\bf v}^{(k)} \} $ leads to the filters satisfying the relations Eqs. (\ref{Eq10}) and (\ref{Eq11}). In the SSA,  the specific $K$-dimensional vectors $ {\bf v} $ are adopted which maximize the norm of the $L$-dimensional vector $X{\bf v}$ preserving the orthonormality of the vectors $ {\bf v} $, which leads to the eigenvalue equation Eq. (\ref{Eq3}). For the eigenvalues and the eigenvectors for the lagged-covariance matrix $C$, the following relation holds
\begin{equation}
\lambda_k={\bf v}^{(k)T}X^TX{\bf v}^{(k)}= \sum_\alpha |{\hat v}_\alpha^{(k)}|^2|{\hat x}_\alpha|^2  \ \ .
\label{Eq12}
\end{equation}
The SSA can be interpreted as an algorithm to determine the orthonormal $K$-dimensional vectors ${\bf v}^{(k)}$ which maximize the quantity  $\sum_\alpha |{\hat v}_\alpha|^2|{\hat x}_\alpha|^2$ preserving the orthonormality of the vectors {\bf v}.  In this sense, the the SSA can be understood as the principal component analysis applied to the trajectory matrix constructed from the lagged vectors. The orthogonality of the eigenvectors are preserved as the phase content for  $\hat{v}_\alpha^{(k)}$, and then, it is lost for the filters $|\hat{v}_\alpha^{(k)}|^2$. Instead of loosing the orthogonality, the filters $|\hat{v}_\alpha^{(k)}|^2$ are real valued and lead no phase shift or distortion for the time series [1,9,10,12].

The SSA decomposition of the time series can simply be expressed as
\begin{equation}
\hat{x}_\alpha = (|\hat{v}^{(1)}_\alpha|^2 \hat{x}_\alpha + |\hat{v}^{(2)}_\alpha|^2 \hat{x}_\alpha + ... + |\hat{v}^{(K)}_\alpha|^2\hat{x}_\alpha )/K  \ . 
\label{Eq13}
\end{equation}
We denote the inverse Fourier transform of the component $ |\hat{v}^{(k)}_\alpha|^2\hat{x}_\alpha$ as $x_n^{(k)}$. Then, the SSA decomposition of the time series is expressed as
\begin{equation}
x_n=x^{(1)}_n+x^{(2)}_n+ \cdot \cdot \cdot + x^{(K)}_n \ \ .
\label{Eq14}
\end{equation}
For later convenience, we denote the filters as
\begin{equation}
F^{(k)}_\alpha=|\hat{v}^{(k)}_\alpha |^2 \ \ ,
\label{Eq15}
\end{equation}
and the Eqs. (\ref{Eq10}) and (\ref{Eq11}) can be rewritten as
\begin{equation}
\sum_\alpha F_\alpha^{(k)}/L = 1 \ \ ,
\label{Eq16}
\end{equation}
and
\begin{equation}
\sum_k F_\alpha^{(k)}/K=1 \ \ .
\label{Eq17}
\end{equation}
From the filtering viewpoint, the SSA algorithm can be seen as the optimal generation of the filter set satisfying the relations Eqs. (\ref{Eq10}) and (\ref{Eq11}), and the time-series decomposition can be done as in Eq.(\ref{Eq13}).  
From the completeness relation Eq. (\ref{Eq17}),  we can see that the decomposition of the power spectrum of the time series $|\hat{x}_\alpha|^2$ is possible as
\begin{equation}
|\hat{x}_\alpha|^2={1 \over K} \sum_{k=1}^K F^{(k)}_\alpha |\hat{x}_\alpha|^2 \ \ .
\label{Eq18}
\end{equation}
Through the above decomposition of power spectrum, we can obtain the insights into the spectral structure of SSA decomposition. On the other hand, squared Fourier transform of the SSA decomposition  $x^{(k)}$ in Eq. (\ref{Eq13}) corresponds to $[\ |\hat{v}^{(k)}_\alpha]^2 \hat{x}_\alpha\ ]^2= |{\hat v}_\alpha^{(i)}|^4|{\hat x}_\alpha|^2$ and their sum does not reproduce the power spectrum of the time series.  In the next section., we will show that the above decomposition can be applied to analyze the details of the SSA decomposition and it can be used to monitor the spectral structure of the decomposition which helps us for the proper choice of the window length parameter $K$.

\section{Example}
To illustrate how the spectral decomposition of the power spectrum described in the last section works, we consider the schematic time series and the daily currency exchange rate EUR/USD.

\begin{figure}[th]
\includegraphics[width=10cm,clip]{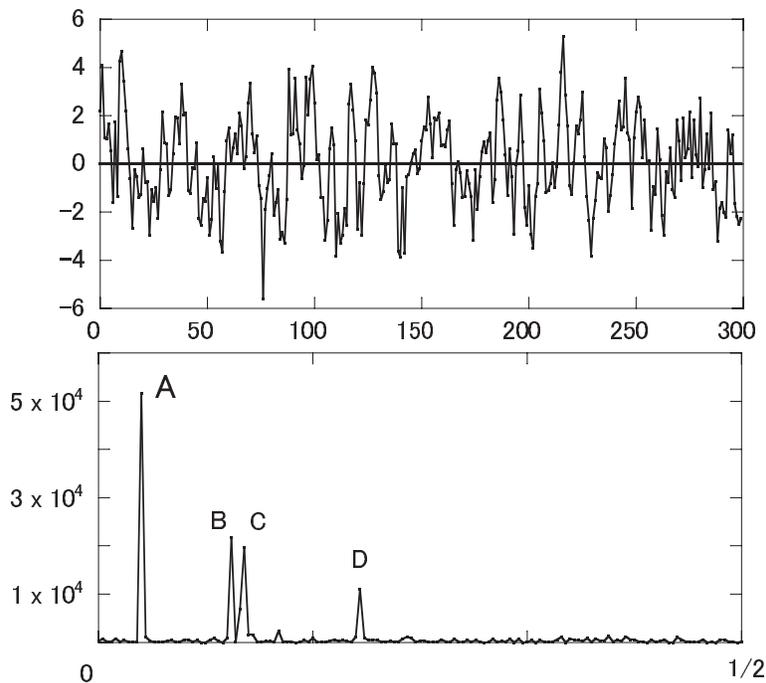}
\centering
\caption{The upper figure is the time series in Eq. (\ref{Eq19}) and the lower is its power spectrum.}
\label{fig1}
\end{figure}

\begin{figure}[th]
\includegraphics[width=9cm,clip]{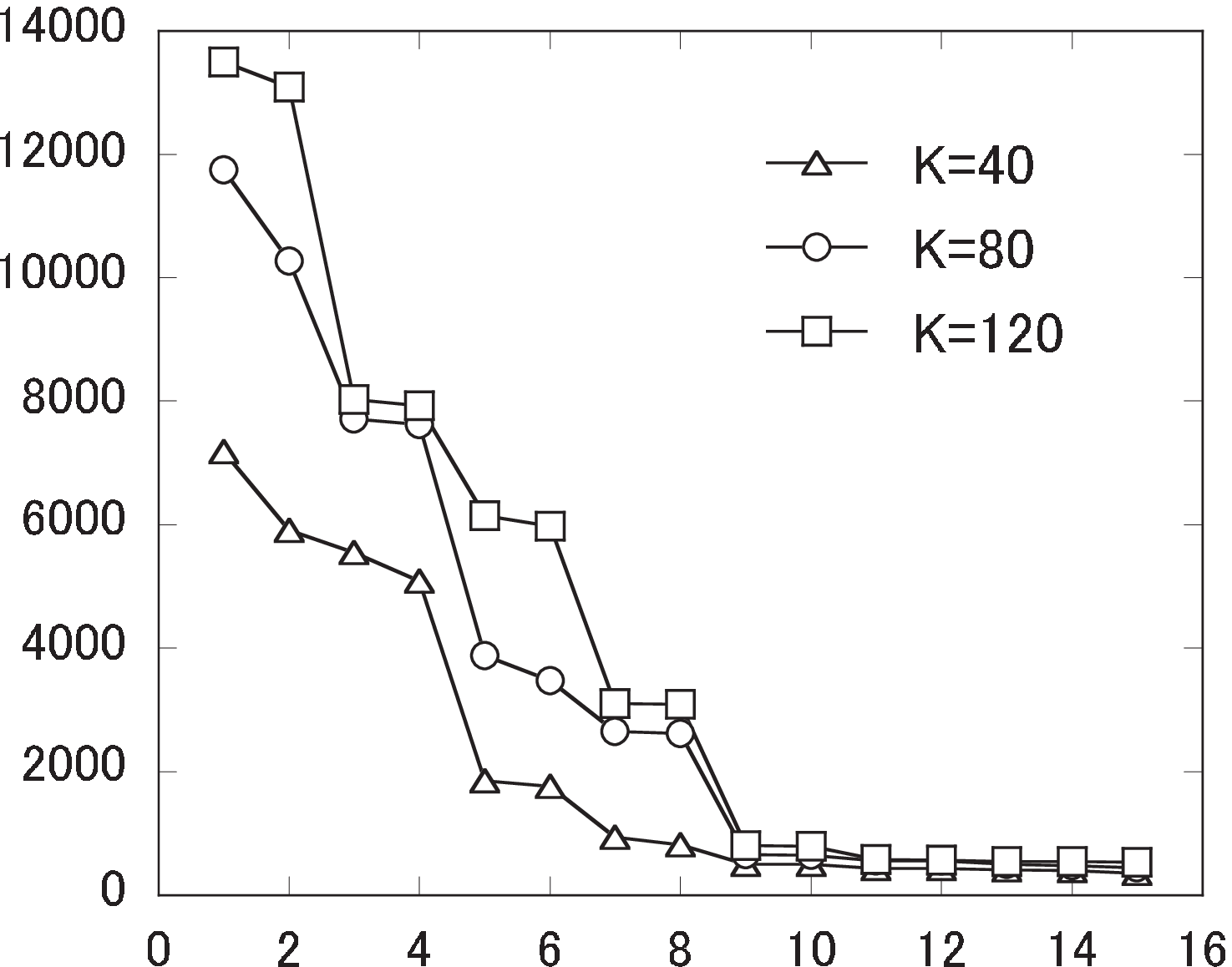}
\centering
\caption{The larger 15 eigenvalues of the lagged-covariance matrix for the window lengths $K=40, 80$ and $120$, respectively. We can see that the eigenvalues are paired for larger $K$, corresponding to the harmonic components in the time series. }
\label{fig2}
\end{figure}

\subsection{Schematic model}
First, we adopt the simple time series which is the superposition of the harmonic components with additive noise,
\begin{equation}
 x_n  =  \sum_{i=1}^4 a_i {\rm sin}(2 \pi n/b_i)+\epsilon_n \ \ ,
\label{Eq19}
\end{equation}
\noindent
where $N=300$ and $\epsilon_n$ the Gaussian noise $N(0,1)$. We adopt the parameters: $a_1=1.44, \   a_2=1.04, \  a_3=1.04, \  a_4=0.8; \ b_1=29.9, \ b_2=9.7, \ b_3=8.9 \ {\rm and} \ b_4=4.9$. The above time series Eq. (\ref{Eq19}) and its power spectrum are shown in Fig. 1. 
Corresponding to the harmonic components, there are four peaks A$\sim$D in the power spectrum. As seen, the second B and the third C peaks are close and have similar strengths. In such cases, these peaks are mixed with each other and are difficult to separate in SSA as is discussed in Ref. [3].   To see the dependence on the choice of the window length, we adopt the values $K=40, \ 80$ and $120$. The larger 15 eigenvalues are shown in Fig. 2. It is known, for the case with larger window lengths, each harmonic component corresponds to the degenerate eigenvalue pair [3,4] and the corresponding eigenvectors are approximately the sinusoid with the relative phase shift $\pi/2$ [3,13]. 
As seen, paired degeneracy of the eigenvalue becomes to be apparent for larger window lengths.  For, $K=120$, we can see clear four degenerate pairs.  
 Since the filters are defined as the squared Fourier transform of the eigenvector of the lagged-covariance matrix as in Eq. (\ref{Eq3}), the Fourier transform $\hat{v}_\alpha^{(k)}$ of the eigenvector for the degenerate pair differ only in phase and leads to approximately the same filters $F^{(k)}$.  These filters extract the corresponding harmonic components from the time series. 
For all cases $K=40, \ 80$ and $120$, the peaks A and D are clearly filtered separately since these are isolated and the strengths are different from the other prominent spectral peaks. 
For the two peaks B and C, however, they are close with each other and the strengths are also similar.  Though we can see the clear pairing of the eigenvalues for $K=120$ in Fig. 2, it is not clear to what extent these peaks are separated in the SSA decomposition.  To see this, we have shown in Fig. 3, the decomposition of the power spectrum in Eq. (\ref{Eq18}).  Since the filter pairs $(F^{(1)}, F^{(2)})$ and $(F^{(7)}, F^{(8)})$ clearly extract the peaks A and D, respectively, these are not shown in the figures. The components $F^{(3)}_\alpha|\hat{x}_\alpha|^2\sim F^{(6)}_\alpha|\hat{x}_\alpha|^2$ are shown for the window lengths $K=40, \ 80$ and $120$, respectively.  As seen, the filtered strengths are fragmented and the peak D component are mixed for smaller window length $K=40$.   For $K=80$, the component of the peak D almost disappears and the four filters $F^{(3)}_\alpha \sim F^{(6)}_\alpha$ extract both B and C components.

\begin{figure}[th]
\includegraphics[width=14cm,clip]{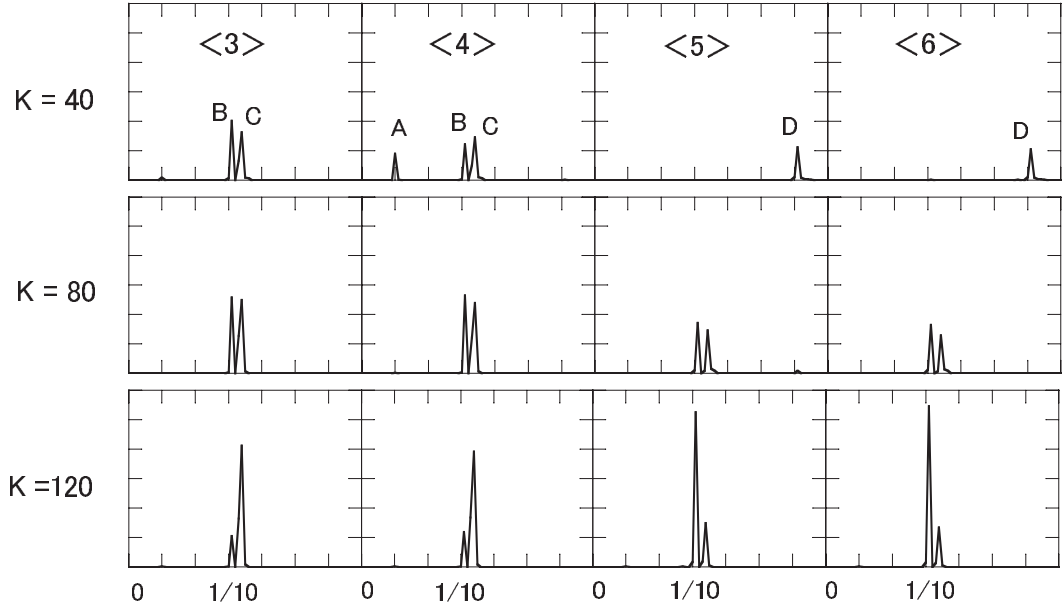}
\centering
\caption{The spectral decompositions $F^{(3)}_\alpha|\hat{x}_\alpha|^2\sim F^{(6)}_\alpha|\hat{x}_\alpha|^2$ are shown for the window lengths $K=40, \ 80$ and $120$, respectively. From left to right, the figures represent the spectrum strengths $F_\alpha^{(k)}|\hat{x}_\alpha|^2$ $(k=3 \sim 6)$ denoted as $<k>$. We take the same scale for all figures. }
\label{fig3}
\end{figure}

\begin{figure}[th]
\includegraphics[width=8cm,clip]{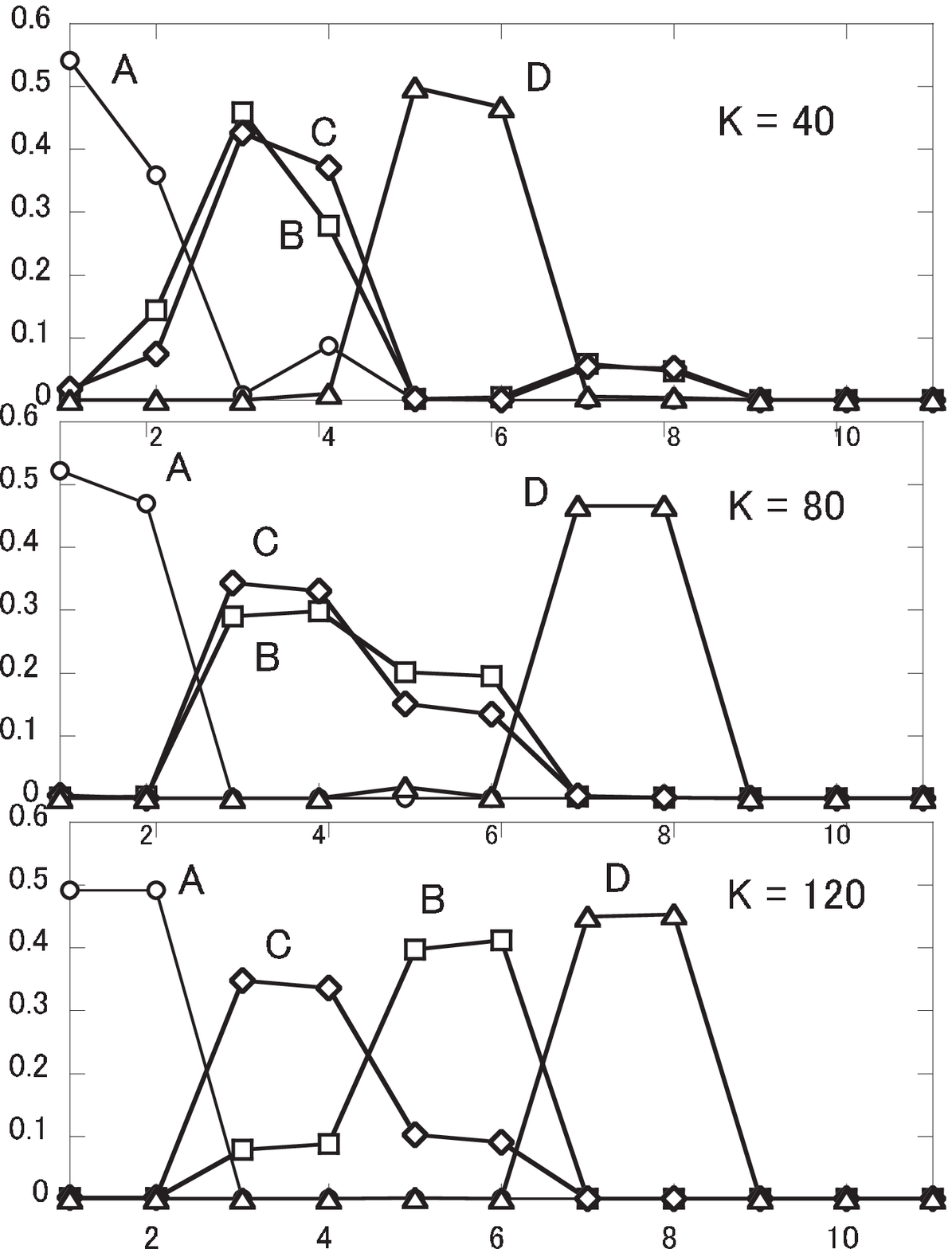}
\centering
\caption{The relative strengths $R_i^{(k)}$ ($k=1 \sim 11$) are shown for the spectral peaks $i$=A,B,C and D for the window lengths 
 $K=40, 80$ and $120$, respectively. }
\label{fig4}
\end{figure}

For larger window length $K=120$, which is close to the half of the length of the time series, the filters $F^{(3)}_\alpha$ and $F^{(4)}_\alpha$  extract mainly the component C with small admixture of B component, while the filters $F^{(5)}_\alpha$ and $F^{(6)}_\alpha$ extract the component B with small admixture of C component. The vector components for the eigenvectors ${\bf v}^{(1)}$ and ${\bf v}^{(2)}$ are the sinusoid with $\pi /2$ phase shift. 
The components of the paired eigenvectors  $({\bf v}^{(3)}, {\bf v}^{(4)})$ or $({\bf v}^{(5)}, {\bf v}^{(6)})$ are the amplitude modulated harmonics with relative phase shift $\pi/2$, representing the mixing of B and C components. These structure is easy to see from the decomposition of the power spectrum.  Instead of this, we can calculate the squared Fourier transform of the decomposition $x^{(k)}$ as $|F^{(k)}_\alpha \hat{x}_\alpha|^2$ but these quantities do not be summed up to reproduce the total power spectrum.  The decomposition of the power spectrum has an advantage that the sum with respect to $\alpha$ corresponds to the eigenvalues as in Eq. (\ref{Eq12}) and are summed up to reproduce the power spectrum as in Eq. (\ref{Eq18}). 

  In order to see the details of the distribution of the spectral decomposition, we have calculated the strength for individual spectral peaks as

\begin{equation}
I_i^{(k)} = \sum_{\alpha \in [{\rm neighbor \ of  \ the \ peak }\ i] } F_\alpha^{(k)}|\hat{x}_\alpha|^2 \ \ \ \ \ \ \ i={\rm A,B,C,D} \ \ ,
\label{Eq20}
\end{equation}
and define the relative strengths $R^{(k)}_i$ as

\begin{equation}
R^{(k)}_i =I_i^{(k)} / \sum_{j=1}^K I_i^{(j)} \ \ .
\label{Eq21}
\end{equation}
The results of the calculated relative strengths $R^{(k)}_i$ are shown in Fig. 4. As is expected, the strengths coming from the paired filters dominate for the isolated peaks A and D for almost all values $K$ considered here. For $K=40$, the peak D component is extracted with the filters $F^{(5)}$ and $F^{(6)}$, while for larger window lengths, it is extracted with the filters $F^{(7)}$ and $F^{(8)}$.  For the peaks B and C, the strength distribution depends on the choice of the window length $K$ as seen in the Fig. 4. For smaller values of $K$,  two peaks B and C are considerably mixed with each other.  For larger window length  $K$, the peaks B and C are almost separated with the paired filters but still remains the small admixture of the components B and C.  Thus, the SSA decomposition $(x^{(1)}, x^{(2)})$ and $(x^{(7)}, x^{(8)})$ are the pure harmonics with lower and higher frequencies, respectively, while the decompositions $x^{(3)} \sim x^{(6)}$ are the admixture of two harmonics.      

For the case that the spectral peaks are well separated, the smaller window length is sufficient to separate the peaks of the spectrum and the results of the decomposition does not strongly dependent on the choice of the window length $K$. For the cases, where the spectral peaks are close or even overlap with each other, the Fourier peaks are mixed and also the SSA decomposition is sensitive to the choice of the window length parameter [3,4]. Thus, for the time series with the complex spectral structure, we should be careful for the choice of the window length parameter. For the proper SSA decomposition, it is helpful to monitor the spectral content of the decomposition and its dependence on the window length $K$ by examining the power-spectrum decomposition as described here.

\begin{figure}[th]
\includegraphics[width=9cm,clip]{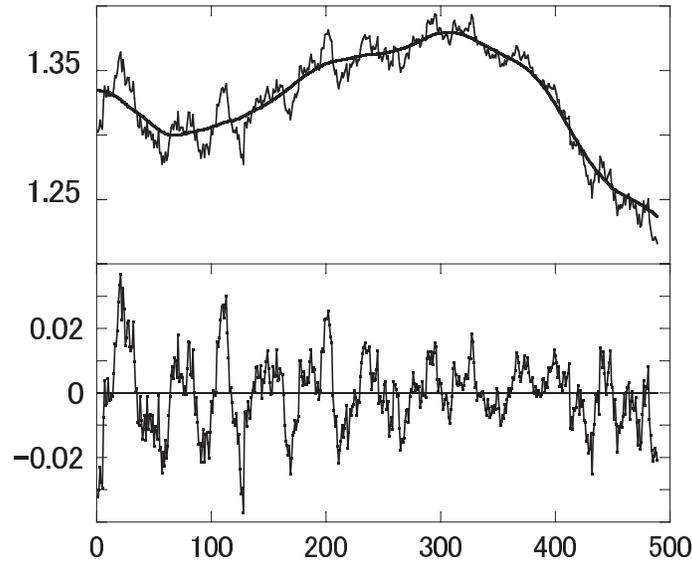}
\centering
\caption{ The daily currency exchange rate EUR/USD for the period 2013-2014. The  number of data points is $N=489$. The upper figure is the original daily rate and the first component of the SSA decomposition with the window length $K$=40. The lower graph is the residual time series. }
\label{fig5}
\end{figure}

\begin{figure}[th]
\includegraphics[width=9cm,clip]{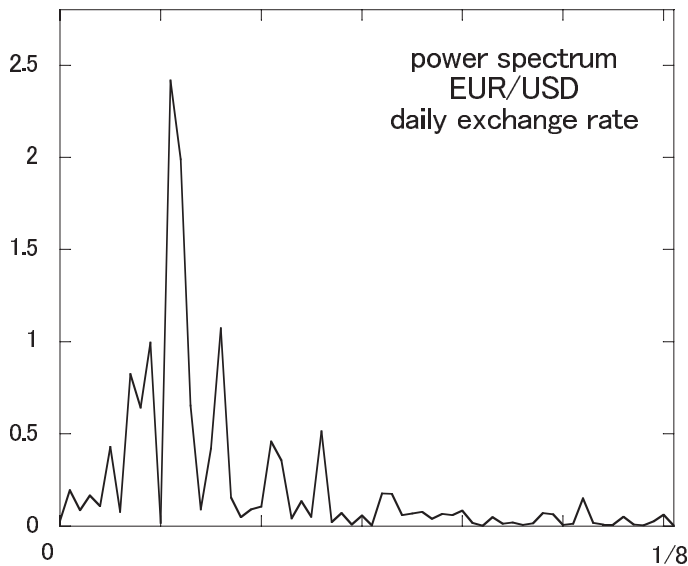}
\centering
\caption{ The power spectrum of the residual components of the currency exchange rate EUR/USD  shown as the lower graph in Fig. 5. }
\label{fig6}
\end{figure}

\begin{figure}[th]
\includegraphics[width=8cm,clip]{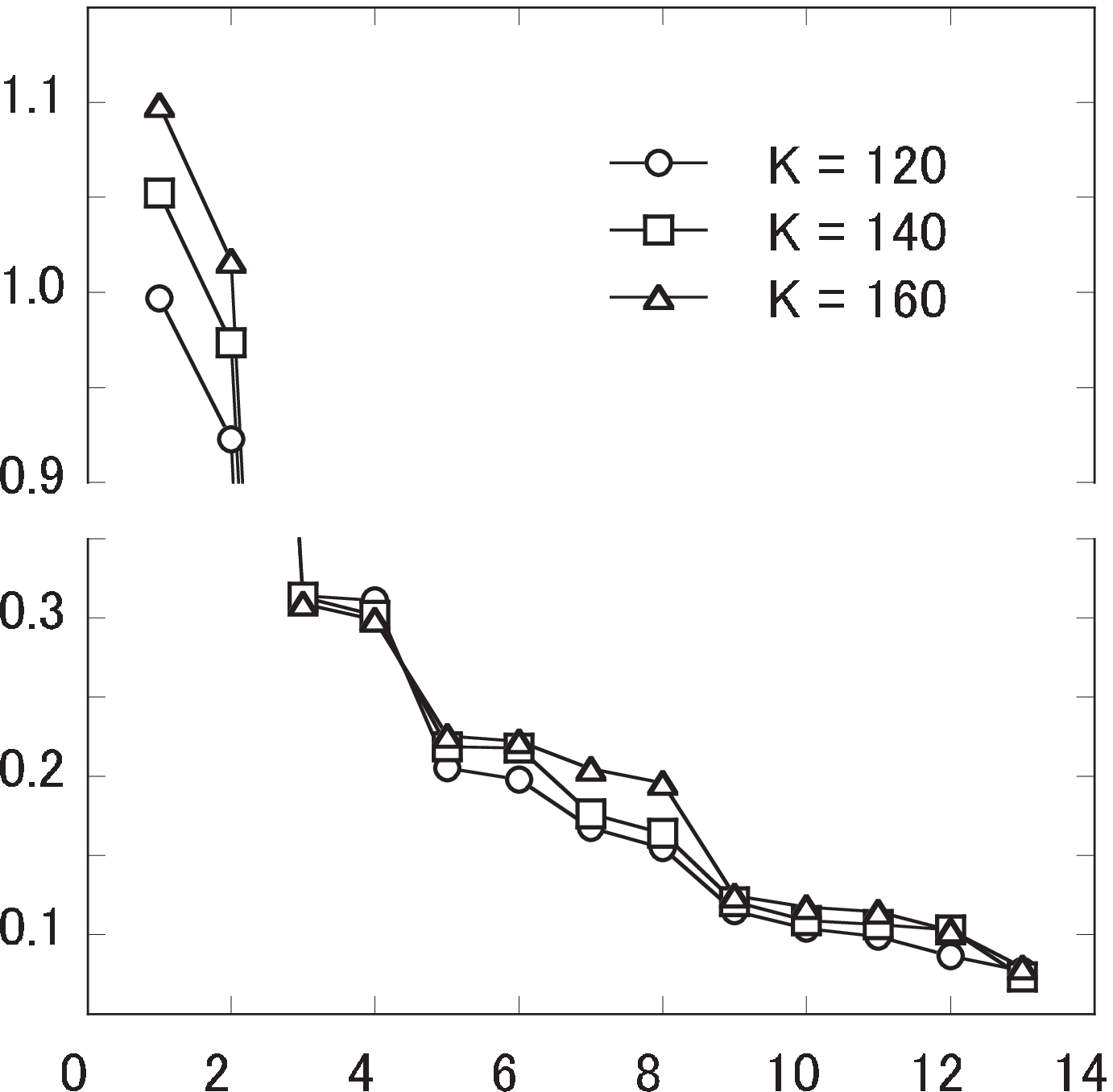}
\centering
\caption{The larger 13 eigenvalues of the lagged-covariance matrix for the residual components with the window lengths $K$=120, 160 and 200, respectively. The first and the second eigenvalues $\lambda_1$ and $\lambda_2$ dominate. }
\label{fig7}
\end{figure}

\subsection{Daily currency exchange rate EUR/USD} 
Next, we have applied the spectral decomposition for the analysis of the daily currency exchange rate Euro vs U.S. dollar for the period 2013-2014. The data consist of $N$=489 points.  The trend component of the series is separated by subtracting the first component of the SSA decomposition with the window length $K=40$, which is called as the sequential SSA [3].  The original exchange rate and the first component of the SSA decomposition are shown in the upper graph in Fig. 5. The residual component is also shown as the lower graph, and its power spectrum is shown in Fig. 6. For the spectrum of the residual series, there are several peaks and are close with each other. Then, in order to see to what extent these harmonic components are separated, we have carried out the SSA decomposition with somewhat large window length parameters $K=140, \ 160$ and 200. The eigenvalues $\lambda_1\sim \lambda_{13}$ are shown in Fig. 7.

\begin{figure}[!htb]
\includegraphics[width=7cm,clip]{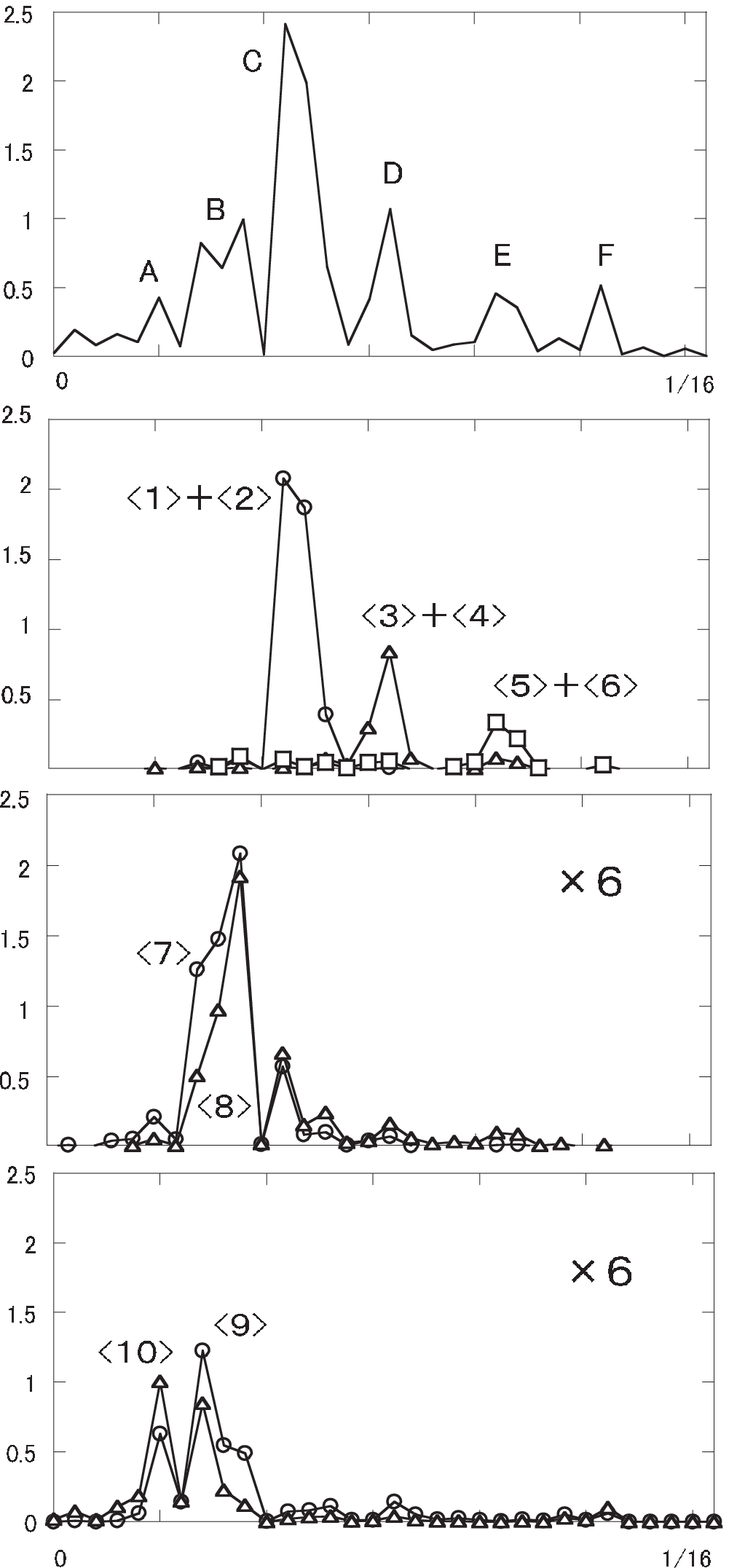}
\centering
\caption{The spectral decompositions $F^{(1)}_\alpha|\hat{x}_\alpha|^2\sim F^{(10)}_\alpha|\hat{x}_\alpha|^2$ are shown for the window length $K=140$.
The individual components $F^{(k)}_\alpha|\hat{x}_\alpha|^2$ are abbreviated as $<k>$. For the lower two graphs, a factor 6 is multiplied. }
\label{fig8}
\end{figure}

\begin{figure}[!htb]
\includegraphics[width=12cm,clip]{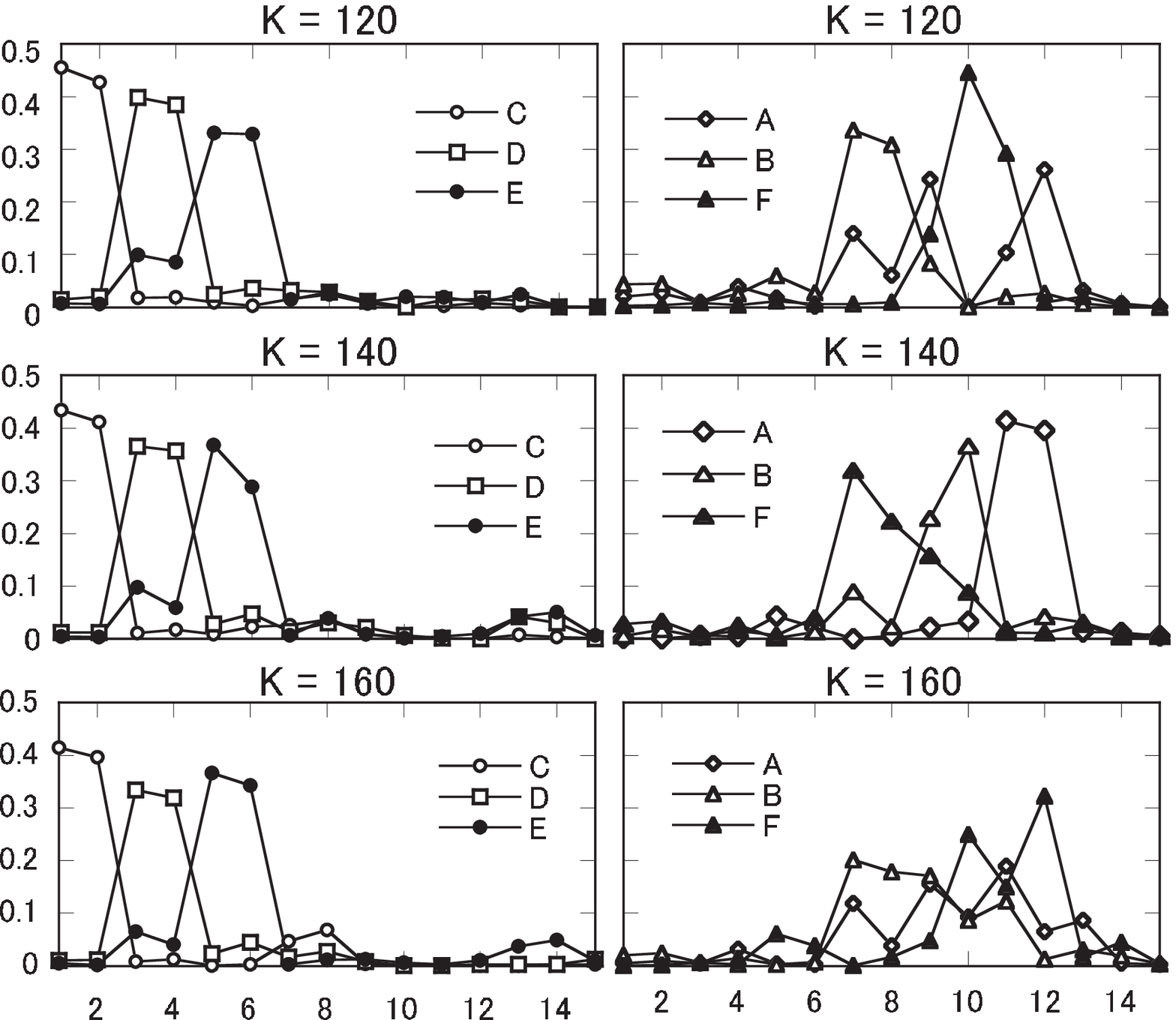}
\centering
\caption{The relative strengths $R_i^{(k)}$ ($k=1 \sim 15$) are shown for the peaks $i$=A $\sim$ F with the window lengths $K=120, \ 140 $ and $160$. }
\label{fig9}
\end{figure}

 In this case, the eigenvalues $(\lambda_1,\lambda_2)$ and $(\lambda_3,\lambda_4)$ are paired corresponding to the harmonic components in the series.  The eigenvalues $(\lambda_5,\lambda_6)$ seems to be paired but are close with the values for the pair $(\lambda_7,\lambda_8)$ and these eigenvalues do not exhibit isolated pairing. There are five or six prominent peaks in the power spectrum and are distinguished by the attached indices A-F as in the upper graph of Fig. 8. In the lower three graphs of Fig. 8, we have shown the decomposition of the power spectrum for the case $K=140$. Since the extracted components with pair of filters $(F^{(1)},F^{(2)})$, $(F^{(3)},F^{(4)})$ and $(F^{(5)},F^{(6)})$ are almost the same and we cannot distinguish them in the figure, we have shown the sum of these components $(F^{(1)}_\alpha+F^{(2)}_\alpha)|\hat{x}_\alpha|^2$,  $(F^{(3)}_\alpha+F^{(4)}_\alpha) | \hat{x}_\alpha|^2$ and  $(F^{(5)}_\alpha+F^{(6)}_\alpha) |\hat{x}_\alpha|^2$ in these figures.  The pair of filters $(F^{(1)},F^{(2)})$ corresponds to the largest peaks C, while the $(F^{(3)},F^{(4)})$ and $(F^{(5)},F^{(6)})$ pairs extract the peak D and E components, respectively. The broader peak B consists of the components $\lambda_7$ and $\lambda_8$, but in this case, there is moderate admixture of dominant C component. The peaks A and B are close with each other and the strengths are the mixture of  $\lambda_9$ and $\lambda_{10}$ components.  In order to see the dependence on the choice of the window length $K$, we have calculated the relative strengths in Eq. (\ref{Eq21}) for the six peaks A-F and the results are shown in Fig. 9 for $K=120,  \ 140$ and $160$, respectively.
For the dominant harmonic components C, D and E, the results do not strongly depend on the choice of window length $K$. For the weak components A, B and F, the details of the filtering are considerably dependent on the choice of the window length $K$. For the separation of the harmonic components, window length around $K=140$ seems to be appropriate.  

In the SSA, the filters $F^{(k)}_\alpha$ are optimally constructed from the time series.  We can say that the SSA algorithm is the soft partitioning of the Fourier spectrum with the normalized filters. Roughly speaking, the average widths of these filters in Fourier space gradually decreases with the increase of the window length $K$ extracting the finer structure of the spectrum, but the structure of these filters are strongly dependent on the power spectrum of the time series itself.  Generally, for dominant eigenvalues, corresponding filter $F^{(k)}_\alpha$ has a simple structure of prominent peak and oscillating tails similar to the squared sinc function. For non-dominant eigenvalues, however, the filters have complicated and broader function of $\alpha$.  Thus, the SSA decomposition is considerably dependent on the choice of window length and it is difficult to find the general criterion for the proper choice of the window length $K$. Because of these, the spectrum decomposition with the filters $F^{(k)}$ are useful to monitor the SSA decomposition in detail and adjust the window length in a specific application of the time series.

\section{Summary and conclusions}

 In this paper, based on the filtering interpretation of SSA, the decomposition of the power spectrum for the original time series is shown to be possible and this helps us for  obtaining the detailed insights into the SSA decomposition and it also helps us to choose the proper window length parameter $K$ for a specific application. 
In the filtering interpretation, the SSA algorithm has been seen as the generation of normalized and complete set of adaptive filters and the SSA decomposition of the time series can be done through the operation of these filters to the original data.  The SSA decomposition is the soft and the optimal partitioning of the Fourier space through the adaptively constructed real-valued filters. By using the completeness property of the eigenvectors of the lagged-covariance matrix, not only the time series itself but also the power spectrum can be decomposed into {\it additive} $K$ components with the filters. The decomposition of the spectrum clearly shows the detailed structure of the SSA decomposition.  Roughly speaking, the widths of the filters $F^{(k)}_\alpha$ slowly decrease with the increase of window length $K$. This suggests that the finer structure of the spectrum of the time series can be separated with larger window length. But the structure of the filters are fairly complex and the the dependence on the window length is not so simple as to set any general criterion for the proper choice of the window length $K$.  Besides eigentriple of the SSA decomposition (eigenvalues, right and left singular values of the trajectory matrix), the decomposition of the power spectrum with the filters is useful for monitoring how the SSA algorithm works for the time-series decomposition and for the proper choice of the window length $K$.    \\

\noindent
{\bf Acknowledgment} \\
\noindent
This work is supported by Grant-in-Aid for Scientific Research from the Japan Society for the Promotion of Science; Grant Numbers (24650150) and (26330237).

\end{document}